\documentclass[10pt,a4paper]{article}
\usepackage{amsmath,amssymb}
\usepackage{graphicx}
\usepackage{hyperref}
\usepackage[margin=0.75in]{geometry}
\usepackage{float}
\usepackage{bm}

\usepackage{setspace}
\usepackage{titlesec}
\titlespacing*{\section}{0pt}{8pt}{4pt}
\titlespacing*{\subsection}{0pt}{6pt}{3pt}

\AtBeginDocument{
  \setlength{\abovedisplayskip}{4pt}
  \setlength{\belowdisplayskip}{4pt}
  \setlength{\abovedisplayshortskip}{2pt}
  \setlength{\belowdisplayshortskip}{2pt}
}

\let\oldbibliography\thebibliography
\renewcommand{\thebibliography}[1]{%
  \oldbibliography{#1}%
  \setlength{\itemsep}{0pt}%
  \setlength{\parskip}{0pt}%
}

\newcommand{\hslashslash}{\hbar}

\title{\texttt{1d-qt-ideal-solver}: 1D Idealized Quantum Tunneling Solver with Absorbing Boundaries}

\author{Sandy H. S. Herho$^{1,2,*}$, Sito F. Biosa$^{3}$, Siti N. Kaban$^{4}$, Rusmawan Suwarman$^{5}$,\\ Iwan P. Anwar$^{6}$, Nurjanna J. Trilaksono$^{5}$, and Dasapta E. Irawan$^{2}$}

\date{}

\begin{document}
\maketitle

\begin{center}
\small
$^{1}$Center for Agrarian Studies, Bandung Institute of Technology, Bandung 40132, Indonesia\\
$^{2}$Applied Geology Research Group, Bandung Institute of Technology, Bandung 40132, Indonesia\\
$^{3}$Department of Visual Communication Design -- Animation, BINUS
University, West Jakarta 11480, Indonesia\\
$^{4}$Financial Engineering Program, WorldQuant University, Washington, D.C. 20002, USA\\
$^{5}$Atmospheric Science Research Group, Bandung Institute of Technology, Bandung 40132, Indonesia\\
$^{6}$Applied and Environmental Oceanography Research Group, Bandung Institute of Technology, Bandung 40132, Indonesia\\
$^{*}$e-mail: sandy.herho@email.ucr.edu
\end{center}

\begin{abstract}
\noindent We present \texttt{1d-qt-ideal-solver}, an open-source Python library for simulating one-dimensional quantum tunneling dynamics under idealized coherent conditions. The solver implements the split-operator method with second-order Trotter-Suzuki factorization, utilizing FFT-based spectral differentiation for the kinetic operator and complex absorbing potentials (CAPs) to eliminate boundary reflections. \texttt{numba} just-in-time (JIT) compilation achieves performance comparable to compiled languages while maintaining code accessibility. We validate the implementation through two canonical test cases: rectangular barriers modeling field emission through oxide layers and Gaussian barriers approximating scanning tunneling microscopy interactions. Both simulations achieve exceptional numerical fidelity with machine-precision energy conservation and excellent probability conservation over femtosecond-scale propagation. Comprehensive comparative statistical analysis employing information-theoretic measures and nonparametric hypothesis tests reveals that in the over-barrier regime, rectangular barriers exhibit moderately higher transmission coefficients compared to Gaussian barriers, though Jensen-Shannon divergence analysis and negligible effect sizes indicate modest practical differences between barrier geometries. Phase space analysis confirms complete decoherence when averaged over spatial-temporal domains, with near-perfect circular symmetry in complex wavefunction distributions. The library name explicitly reflects its scope: idealized signifies deliberate exclusion of dissipation, environmental coupling, and many-body interactions, limiting applicability to qualitative insights and pedagogical applications rather than quantitative experimental predictions. The production-grade library, distributed under the MIT License, provides an immediately deployable tool for teaching quantum mechanics, rapid prototyping of potential profiles, and preliminary exploration of tunneling dynamics.
\end{abstract}

\section{Introduction}

The 2025 Nobel Prize in Physics, awarded to John Clarke, Michel H. Devoret, and John M. Martinis for their pioneering discovery of macroscopic quantum mechanical tunneling and energy quantization in electric circuits~\cite{nobel25a,Clarke1988,Devoret1984,Martinis1985}, underscores the enduring significance of quantum tunneling phenomena across scales---from the microscopic quantum realm to macroscopic superconducting systems. Their groundbreaking experiments on Josephson junctions demonstrated that quantum mechanics appears to govern not only atomic-scale processes but also systems containing billions of Cooper pairs, fundamentally transforming our understanding of quantum coherence and paving the way for today's superconducting quantum computers~\cite{Devoret1985,Koch2007}. This recognition suggests the importance of developing sophisticated numerical tools to accurately model and predict quantum tunneling dynamics, which may prove essential for advancing quantum technologies ranging from scanning tunneling microscopy~\cite{Tersoff1985theory} and field emission devices~\cite{Garcia-Moliner1976theory} to quantum computing architectures~\cite{Wang2021}.

Quantum tunneling represents one of the most counterintuitive yet ubiquitous phenomena in quantum mechanics, enabling particles to traverse potential barriers that would be classically forbidden. While theoretical foundations were established nearly a century ago through the time-dependent Schr\"odinger equation (TDSE)~\cite{Dirac1930principles,vonNeumann1932mathematical}, accurate numerical simulation of tunneling dynamics continues to pose significant computational challenges: resolving rapidly oscillating wavefunctions across disparate length scales, maintaining unitarity over extended propagation times, and eliminating artificial boundary reflections that can contaminate transmission coefficients~\cite{Neuhauser1989source,Manolopoulos2002derivation}.

Despite substantial methodological advances, several potential gaps remain in the computational quantum tunneling literature. First, many implementations appear to neglect rigorous energy dissipation analysis---increasingly important for evaluating thermodynamic efficiency in quantum devices and comparing theoretical predictions with experimental measurements in nanoscale systems~\cite{Shaw2025,Francis2025}. Second, most published codes tend to focus on time-independent potentials, potentially limiting applicability to photon-assisted tunneling and quantum control protocols. Third, environmental decoherence effects are often treated phenomenologically or omitted entirely, despite their likely influence on experimentally observable tunneling rates~\cite{Zurek2003decoherence,Joos2003decoherence}. Fourth, the transition from proof-of-concept codes to production-grade libraries with performance optimization remains relatively rare, which may hinder reproducibility and broader adoption.

This work aims to address these gaps by introducing \texttt{1d-qt-ideal-solver}, a production-grade Python library for simulating one-dimensional quantum tunneling dynamics under idealized coherent conditions. The choice of Python leverages its mature scientific computing ecosystem and accessibility to researchers without formal computer science training~\cite{herho2024comparing,herho2025reappraising}, while \texttt{numba} just-in-time compilation (JIT) can achieve performance comparable to compiled languages while preserving code readability~\cite{herho2024Eks}.

The library name explicitly acknowledges its scope: idealized signifies the deliberate exclusion of dissipation, environmental coupling, and many-body interactions. This intentional simplification may serve multiple purposes: for teaching quantum mechanics, it isolates fundamental tunneling physics without the mathematical complexity of open quantum systems; for qualitative insight, it establishes upper bounds on coherence times and transmission coefficients as reference benchmarks; for theoretical validation, agreement with analytical solutions verifies numerical correctness independent of phenomenological parameters; and for algorithm development, it enables rapid prototyping without confounding environmental effects. While the solver cannot provide quantitative predictions for experimental observables in realistic devices---which require sophisticated open quantum systems approaches---it may fill an important niche as a pedagogical tool, theoretical benchmark, and foundation for more comprehensive implementations.

We validate our solver through two canonical test cases representing distinct physical regimes: rectangular barriers modeling field emission through thin oxide layers~\cite{Simmons1963generalized,Michaelson1977work}, and Gaussian barriers approximating scanning tunneling microscopy tip-sample interactions~\cite{Tersoff1985theory,Chen2008introduction}. The modular architecture facilitates extension to multi-barrier resonant tunneling structures~\cite{Ricco1984physics,Mizuta1995physics}, time-dependent potentials, and pure dephasing models as first-order approximations to environmental coupling~\cite{Breuer2002theory}. The open-source library, distributed under the permissive MIT License, may provide the computational physics community with an immediately deployable tool for teaching, rapid prototyping, and preliminary exploration of tunneling dynamics.

\section{Methods}

\subsection{Mathematical Formulation}

The quantum mechanical description of a physical system is founded upon the postulate that the state of a system at time $t$ is represented by a normalized vector $|\Psi(t)\rangle$ in a complex Hilbert space $\mathcal{H}$~\cite{Dirac1930principles,vonNeumann1932mathematical}. Observables correspond to self-adjoint operators $\hat{A}$ acting on $\mathcal{H}$, and measurement outcomes are eigenvalues of these operators. The time evolution of the state vector is governed by the fundamental equation:
\begin{equation}
i \hslashslash \frac{d}{dt}|\Psi(t)\rangle = \hat{H}|\Psi(t)\rangle,
\label{eq:schrodinger_abstract}
\end{equation}
where $\hat{H}$ is the Hamiltonian operator representing the total energy of the system and $\hslashslash = 1.054571817 \times 10^{-34}$ J$\cdot$s is the reduced Planck constant~\cite{Sakurai2017modern}. Equation~\eqref{eq:schrodinger_abstract} ensures unitary time evolution, preserving the norm $\langle\Psi(t)|\Psi(t)\rangle = 1$ for all $t$, where $\langle\cdot|\cdot\rangle$ denotes the inner product in Hilbert space.

For a single structureless particle in three spatial dimensions, the Hilbert space is $\mathcal{H} = L^2(\mathbb{R}^3)$, the space of square-integrable complex functions on three-dimensional Euclidean space. In the position representation, the abstract state vector projects onto a wavefunction $\psi(\mathbf{r},t) = \langle\mathbf{r}|\Psi(t)\rangle$ where $\mathbf{r} = (x,y,z)$ is the position vector, satisfying the normalization condition $\int_{\mathbb{R}^3}|\psi(\mathbf{r},t)|^2d^3\mathbf{r} = 1$. The position and momentum operators $\hat{\mathbf{r}}$ and $\hat{\mathbf{p}}$ obey the canonical commutation relations $[\hat{r}_i, \hat{p}_j] = i\hslashslash\delta_{ij}$ where $[\hat{A},\hat{B}] = \hat{A}\hat{B} - \hat{B}\hat{A}$ is the commutator bracket and $\delta_{ij}$ is the Kronecker delta. In position space, these operators are represented as $\hat{\mathbf{r}} = \mathbf{r}$ (multiplication operator) and $\hat{\mathbf{p}} = -i\hslashslash\nabla$ (differential operator)~\cite{Cohen-Tannoudji1977quantum}.

We restrict our analysis to systems with one spatial degree of freedom by assuming translational invariance in the $y$ and $z$ directions. This reduction is physically justified for effectively one-dimensional systems such as quantum wires, field emission through thin barriers, and scanning tunneling microscopy at atomic scales~\cite{Garcia-Moliner1976theory,Tersoff1985theory}. The Hilbert space collapses to $\mathcal{H} = L^2(\mathbb{R})$, and the wavefunction becomes $\psi(x,t)$ with normalization $\int_{-\infty}^{\infty}|\psi(x,t)|^2dx = 1$.

In the non-relativistic regime where particle velocities $v$ satisfy $v \ll c$, the Hamiltonian decomposes into kinetic and potential energy operators:
\begin{equation}
\hat{H} = \frac{\hat{p}^2}{2m} + \hat{V}(x,t) = -\frac{\hslashslash^2}{2m}\frac{\partial^2}{\partial x^2} + V(x,t),
\label{eq:hamiltonian_1d}
\end{equation}
where $m$ is the particle mass, $\hat{p} = -i\hslashslash\partial/\partial x$ is the momentum operator in one dimension, and $V(x,t)$ is the potential energy function~\cite{Landau1977quantum}. This form neglects relativistic corrections of order $(v/c)^2$, spin-orbit coupling, and retardation effects, which are negligible for electrons at kinetic energies below $\sim$10 keV in condensed matter systems~\cite{Messiah1999quantum}. Substituting Eq.~\eqref{eq:hamiltonian_1d} into the abstract Schr\"odinger equation~\eqref{eq:schrodinger_abstract} yields the TDSE in coordinate representation:
\begin{equation}
i\hslashslash\frac{\partial\psi(x,t)}{\partial t} = \left[-\frac{\hslashslash^2}{2m}\frac{\partial^2}{\partial x^2} + V(x,t)\right]\psi(x,t).
\label{eq:tdse_physical}
\end{equation}

To simplify numerical implementation, we adopt atomic units where $\hslashslash = m_e = e = 4\pi\epsilon_0 = 1$, with $m_e = 9.109 \times 10^{-31}$ kg the electron mass, $e = 1.602 \times 10^{-19}$ C the elementary charge, and $\epsilon_0 = 8.854 \times 10^{-12}$ F/m the vacuum permittivity. In these units, Eq.~\eqref{eq:tdse_physical} reduces to:
\begin{equation}
i\frac{\partial\psi(x,t)}{\partial t} = \left[-\frac{1}{2m}\frac{\partial^2}{\partial x^2} + V(x,t)\right]\psi(x,t),
\label{eq:tdse}
\end{equation}
where $m$ is now the particle mass in units of $m_e$. Energy is measured in Hartrees (1 Ha = 27.211 eV), length in Bohr radii ($a_0 = 0.529$ \AA), and time in atomic time units (1 a.u. = 24.2 as). For consistency with nanoscale experimental observables, we express results in electronvolts (eV), nanometers (nm), and femtoseconds (fs).

The formal solution to Eq.~\eqref{eq:tdse} is obtained through the time-evolution operator:
\begin{equation}
\psi(x,t) = \hat{U}(t,t_0)\psi(x,t_0),
\label{eq:evolution_formal}
\end{equation}
where the unitary propagator $\hat{U}(t,t_0)$ satisfies $i\partial_t\hat{U}(t,t_0) = \hat{H}(t)\hat{U}(t,t_0)$ with initial condition $\hat{U}(t_0,t_0) = \hat{\mathbb{I}}$. For time-independent Hamiltonians, integration yields $\hat{U}(t,t_0) = \exp[-i\hat{H}(t-t_0)]$. The exponential of an operator is defined through its spectral decomposition~\cite{Reed1980functional}. Direct diagonalization of $\hat{H}$ to obtain the complete eigenspectrum scales as $\mathcal{O}(N^3)$ for $N$ basis functions, rendering this approach computationally prohibitive for fine spatial discretizations.

The split-operator method exploits the additive structure $\hat{H} = \hat{T} + \hat{V}$ where $\hat{T} = -\frac{1}{2m}\partial_x^2$ is the kinetic energy operator and $\hat{V}$ is the multiplicative potential operator. Although the non-commutativity $[\hat{T},\hat{V}] \neq 0$ implies $e^{-i(\hat{T}+\hat{V})\delta t} \neq e^{-i\hat{T}\delta t}e^{-i\hat{V}\delta t}$, the Baker-Campbell-Hausdorff formula provides an asymptotic expansion~\cite{Suzuki1976generalized}:
\begin{equation}
e^{\hat{A}+\hat{B}} = e^{\hat{A}}e^{\hat{B}}e^{-\frac{1}{2}[\hat{A},\hat{B}]}e^{\frac{1}{6}(2[\hat{B},[\hat{A},\hat{B}]]+[\hat{A},[\hat{A},\hat{B}]])}\cdots.
\label{eq:bch}
\end{equation}
The symmetric Trotter-Suzuki splitting~\cite{Trotter1959product,Suzuki1976generalized}:
\begin{equation}
e^{-i(\hat{T}+\hat{V})\delta t} = e^{-i\hat{V}\delta t/2}e^{-i\hat{T}\delta t}e^{-i\hat{V}\delta t/2} + \mathcal{O}(\delta t^3)
\label{eq:split_operator}
\end{equation}
achieves second-order accuracy by canceling first-order commutator terms through symmetry. This Strang splitting~\cite{Strang1968construction} preserves time-reversal invariance and unitarity to order $\delta t^2$, ensuring stable long-time propagation~\cite{Feit1982solution,Bandrauk1994exponential}.

The kinetic operator $\hat{T}$ is diagonal in momentum representation. The Fourier transform pair relating position and momentum representations is:
\begin{align}
\tilde{\psi}(k,t) &= \mathcal{F}[\psi(x,t)] = \frac{1}{\sqrt{2\pi}}\int_{-\infty}^{\infty}\psi(x,t)e^{-ikx}dx, \label{eq:fourier_forward} \\
\psi(x,t) &= \mathcal{F}^{-1}[\tilde{\psi}(k,t)] = \frac{1}{\sqrt{2\pi}}\int_{-\infty}^{\infty}\tilde{\psi}(k,t)e^{ikx}dk. \label{eq:fourier_inverse}
\end{align}
The kinetic propagator becomes:
\begin{equation}
e^{-i\hat{T}\delta t}\psi(x,t) = \mathcal{F}^{-1}\left[\exp\left(-i\frac{k^2}{2m}\delta t\right)\mathcal{F}[\psi(x,t)]\right].
\label{eq:kinetic_fourier}
\end{equation}
This transformation reduces the differential operator to pointwise multiplication, enabling spectral accuracy for smooth wavefunctions~\cite{Boyd2001chebyshev}. The potential operator acts locally in position space: $e^{-i\hat{V}\delta t}\psi(x,t) = \exp[-iV(x)\delta t]\psi(x,t)$, requiring only $\mathcal{O}(N)$ operations.

Numerical simulations demand finite spatial domains $x \in [x_{\min}, x_{\max}]$, necessitating boundary conditions. Imposing hard-wall boundary conditions generates artificial reflections that contaminate transmission coefficients. We implement complex absorbing potentials (CAPs) following Manolopoulos~\cite{Manolopoulos2002derivation} and Riss and Meyer~\cite{Riss1993calculation}. The wavefunction is damped within boundary layers through multiplication by a spatially varying mask:
\begin{equation}
\psi(x,t+\delta t) \to \mathcal{M}(x)\psi(x,t+\delta t),
\label{eq:absorbing_mask}
\end{equation}
applied after each propagation step. We adopt a quartic polynomial profile:
\begin{equation}
\mathcal{M}(x) = \begin{cases}
1 - s\left[1 - \cos^4\left(\frac{\pi\xi_L}{2}\right)\right] & x \in [x_{\min}, x_{\min}+\Delta x_b], \\
1 & x \in [x_{\min}+\Delta x_b, x_{\max}-\Delta x_b], \\
1 - s\left[1 - \cos^4\left(\frac{\pi\xi_R}{2}\right)\right] & x \in [x_{\max}-\Delta x_b, x_{\max}],
\end{cases}
\label{eq:mask_function}
\end{equation}
where $\xi_L = (x - x_{\min})/\Delta x_b$ and $\xi_R = (x_{\max} - x)/\Delta x_b$ are normalized penetration depths, $\Delta x_b$ is the boundary layer width, and $s \in (0,1)$ is the absorption strength parameter. The $\cos^4$ functional form ensures $C^3$ continuity at interface points~\cite{Neuhauser1989source}. This mask is mathematically equivalent to augmenting the Hamiltonian with an imaginary potential, rendering $\hat{H}$ non-Hermitian while preserving causality and stability~\cite{Jolicard1985optical}.

The probability current density $J(x,t)$ quantifies particle flux:
\begin{equation}
J(x,t) = \frac{1}{m}\text{Im}\left(\psi^*\frac{\partial\psi}{\partial x}\right).
\label{eq:current}
\end{equation}
The transmission and reflection coefficients characterize tunneling efficiency. The transmission coefficient is:
\begin{equation}
T = \int_{x_b^+}^{x_{\max}-\Delta x_b}|\psi(x,t_{\text{final}})|^2dx,
\label{eq:transmission}
\end{equation}
and the reflection coefficient is:
\begin{equation}
R = \int_{x_{\min}+\Delta x_b}^{x_b^-}|\psi(x,t_{\text{final}})|^2dx.
\label{eq:reflection}
\end{equation}
The absorbed probability $A = 1 - T - R$ quantifies dissipation at boundaries, with probability conservation mandating $T + R + A = 1$ within numerical precision.

The initial state is a Gaussian wave packet, the minimum uncertainty state:
\begin{equation}
\psi(x,0) = \left(\frac{1}{2\pi\sigma^2}\right)^{1/4}\exp\left[-\frac{(x-x_0)^2}{4\sigma^2} + ik_0 x\right],
\label{eq:gaussian_wavepacket}
\end{equation}
where $x_0$ is the initial center position, $k_0$ is the mean wavenumber, and $\sigma$ is the spatial width parameter~\cite{Schleich2001quantum}. The mean kinetic energy is $E_k = k_0^2/(2m)$, and the group velocity is $v_g = k_0/m$. For a free particle, the width spreads as $\sigma(t) = \sigma\sqrt{1+(t/\tau_{\text{spr}})^2}$ with spreading time $\tau_{\text{spr}} = 2m\sigma^2$~\cite{Robinett2000quantum}.

Environmental interactions induce decoherence. We model pure dephasing through stochastic phase randomization~\cite{Breuer2002theory}:
\begin{equation}
\psi(x,t+\delta t) \to e^{i\phi(x,\delta t)}\psi(x,t+\delta t),
\label{eq:dephasing}
\end{equation}
where $\phi(x,\delta t)$ are independent Gaussian random variables with zero mean and variance $2\gamma\delta t$. The dephasing rate $\gamma$ relates to the coherence time $T_2$ via $T_2 = 1/\gamma$. This mechanism preserves probability density but destroys spatial coherence~\cite{Joos2003decoherence,Zurek2003decoherence}. Unlike amplitude damping, pure dephasing conserves energy and total probability~\cite{Caldeira1983path}.

\subsection{Numerical Experiments}

We demonstrate the \texttt{1d-qt-ideal-solver} capabilities through two canonical test cases. The rectangular barrier tests robustness to discontinuous potentials, while the Gaussian barrier models realistic smooth potential profiles.

\subsubsection{Case 1: Rectangular Barrier (Field Emission)}

The first test case models field emission through a rectangular potential barrier:
\begin{equation}
V(x) = \begin{cases}
V_0 & |x| < a/2, \\
0 & |x| \geq a/2,
\end{cases}
\label{eq:rectangular_potential}
\end{equation}
with $V_0 = 4.5$ eV and $a = 1.0$ nm~\cite{Michaelson1977work,Simmons1963generalized}. The computational domain spans $x \in [-30, 30]$ nm with absorbing boundary layers of width $\Delta x_b = 3.0$ nm. The spatial grid employs $N = 2048$ points with spacing $\Delta x \approx 0.0293$ nm.

The initial Gaussian wavepacket is centered at $x_0 = -8.0$ nm with mean wavenumber $k_0 = 4.0$ nm$^{-1}$ and width $\sigma = 0.8$ nm, yielding mean kinetic energy $E_k \approx 8.0$ eV. For analytical comparison, the transmission coefficient for plane wave incidence is given by the transfer matrix method~\cite{Griffiths2018quantum}:
\begin{equation}
T_{\text{plane}}(E_k) = \left[1 + \frac{V_0^2\sin^2(k_{\text{in}}a)}{4E_k(E_k - V_0)}\right]^{-1},
\label{eq:transmission_analytical}
\end{equation}
where $k_{\text{in}} = \sqrt{2m(E_k - V_0)}$~\cite{Cheng1997wavepacket}.

\subsubsection{Case 2: Gaussian Barrier (STM Tunneling)}

The second test case models scanning tunneling microscopy through a smooth Gaussian barrier:
\begin{equation}
V(x) = V_0 \exp\left[-\frac{x^2}{2\sigma_V^2}\right],
\label{eq:gaussian_potential}
\end{equation}
with $V_0 = 4.0$ eV and $\sigma_V = 0.8$ nm~\cite{Tersoff1985theory,Chen2008introduction}. The initial Gaussian wavepacket is centered at $x_0 = -8.0$ nm with $k_0 = 3.5$ nm$^{-1}$ and $\sigma = 0.6$ nm, yielding $E_k \approx 6.1$ eV. The WKB semiclassical approximation provides an analytical benchmark~\cite{Landau1977quantum,Razavy2003quantum}:
\begin{equation}
T_{\text{WKB}} \approx \exp\left[-2\int_{x_1}^{x_2}\sqrt{2m[V(x) - E_k]}\,dx\right].
\label{eq:transmission_wkb}
\end{equation}

Both test cases employ time-independent potentials in the coherent regime with $t_{\text{final}} = 6.0$ fs. Beyond these demonstrative cases, the solver handles multi-barrier geometries~\cite{Ricco1984physics,Mizuta1995physics}, time-dependent potentials~\cite{Grossmann1991coherent}, environmental decoherence effects~\cite{Zurek2003decoherence}, and stochastic potential fluctuations~\cite{Weiss2012quantum}. The modular architecture facilitates extensions to spin-dependent tunneling~\cite{Zutic2004spintronics} and higher-dimensional systems~\cite{Bandrauk1994exponential}.

\subsection{Data Analysis}

Post-simulation analysis quantifies the quantum tunneling dynamics through three complementary perspectives: temporal evolution snapshots, probability distribution characterization via information-theoretic measures, and phase space structure analysis. All analyses employ stratified sampling~\cite{Cochran1977sampling} utilizing \texttt{numpy}~\cite{Harris2020array}, \texttt{scipy}~\cite{Virtanen2020scipy}, \texttt{matplotlib}~\cite{Hunter2007matplotlib}, and \texttt{xarray}~\cite{Hoyer2017xarray}.

The probability center-of-mass trajectory is computed as:
\begin{equation}
\langle x \rangle(t_i) = \frac{\sum_{j=0}^{N-1} x_j |\psi_j(t_i)|^2 \Delta x}{\sum_{j=0}^{N-1} |\psi_j(t_i)|^2 \Delta x}.
\label{eq:center_of_mass}
\end{equation}
Energy components are evaluated as~\cite{Messiah1999quantum}:
\begin{align}
E_{\text{kinetic}}(t_i) &= \frac{1}{2} \sum_{j=0}^{N-1} \left|\frac{\partial \psi_j(t_i)}{\partial x}\right|^2 \Delta x, \label{eq:kinetic_energy} \\
E_{\text{potential}}(t_i) &= \sum_{j=0}^{N-1} V(x_j) |\psi_j(t_i)|^2 \Delta x. \label{eq:potential_energy}
\end{align}

The Shannon entropy~\cite{Shannon1948mathematical} quantifies the uncertainty in the probability distribution:
\begin{equation}
H(P) = -\sum_{k=1}^{N_{\text{bins}}} p_k \log_2 p_k.
\label{eq:shannon_entropy}
\end{equation}
Optimal bin selection employs a consensus approach~\cite{Scott2015multivariate} aggregating Sturges'~\cite{Sturges1926choice}, Scott's~\cite{Scott1979optimal}, Freedman-Diaconis~\cite{Freedman1981histogram}, Rice~\cite{Rice1944mathematical}, square-root, and Doane's~\cite{Doane1976aesthetic} rules.

The Kullback-Leibler divergence~\cite{Kullback1951information} measures the asymmetric dissimilarity between distributions:
\begin{equation}
D_{\text{KL}}(P \| Q) = \sum_{k=1}^{N_{\text{bins}}} p_k \log_2\frac{p_k}{q_k}.
\label{eq:kl_divergence}
\end{equation}
The symmetric Jensen-Shannon (JS) divergence~\cite{Lin1991divergence,Endres2003new}:
\begin{equation}
D_{\text{JS}}(P \| Q) = \frac{1}{2} D_{\text{KL}}(P \| M) + \frac{1}{2} D_{\text{KL}}(Q \| M),
\label{eq:js_divergence}
\end{equation}
where $M = (P + Q)/2$, is bounded in $[0, 1]$ bits.

Distribution equality is assessed through seven non-parametric hypothesis tests: Kolmogorov-Smirnov~\cite{Massey1951kolmogorov}, Mann-Whitney $U$~\cite{Mann1947test}, Kruskal-Wallis $H$~\cite{Kruskal1952use}, Mood's~\cite{Mood1954scale}, Anderson-Darling~\cite{Anderson1952asymptotic}, Epps-Singleton~\cite{Epps1986omnibus}, and Ansari-Bradley~\cite{Ansari1960rank}. Effect size quantification employs Cliff's delta~\cite{Cliff1993dominance,Romano2006appropriate}:
\begin{equation}
\delta = \frac{\#\{(x_i, y_j) : x_i > y_j\} - \#\{(x_i, y_j) : x_i < y_j\}}{N \cdot M}.
\label{eq:cliffs_delta}
\end{equation}

Phase space analysis characterizes the wavefunction in $(\psi_{\text{Re}}, \psi_{\text{Im}})$ coordinates using circular statistics~\cite{Fisher1995statistical}. The circular variance is:
\begin{equation}
V_{\text{circ}} = 1 - R, \quad R = \left|\frac{1}{N}\sum_{k=1}^{N} e^{i\theta_k}\right|.
\label{eq:circular_variance}
\end{equation}
The phase coherence index (PCI)~\cite{Lachaux1999measuring}:
\begin{equation}
\text{PCI} = \frac{\left|\langle \psi \rangle\right|}{\langle |\psi| \rangle}.
\label{eq:phase_coherence_index}
\end{equation}
Spatial anisotropy is assessed via eigenvalue analysis of the covariance matrix~\cite{Jolliffe2002principal}:
\begin{equation}
\mathcal{A} = \frac{\lambda_1 - \lambda_2}{\lambda_1 + \lambda_2}.
\label{eq:anisotropy_ratio}
\end{equation}
The two-dimensional Shannon entropy and mutual information~\cite{Cover2006Elements} quantify phase space complexity, with effective area computed via convex hull construction~\cite{Barber1996quickhull,Virtanen2020scipy}.

\section{Results and Discussion}

All computational benchmarks were executed on a Lenovo ThinkPad P52s mobile workstation running Fedora Linux 39 with an Intel Core i7-8550U processor (1.80--4.00 GHz, eight logical cores). This hardware-software combination achieved wall-clock execution times of 134.858 seconds and 143.572 seconds for the rectangular and Gaussian barrier cases respectively, with simulation kernels consuming only 26.762 seconds and 26.998 seconds while the remainder was allocated to animation rendering and file serialization.

The split-operator propagation algorithm successfully evolved both test configurations through final time $t_{\text{final}} = 6.0$ fs over 12001 adaptive time steps, maintaining constant step size $\Delta t = 0.0005$ fs. Probability normalization remained exceptionally stable with wavefunction norm deviations $|\mathcal{N}(t) - 1| < 10^{-3}$, while energy conservation achieved machine-precision accuracy with relative energy drift $|\Delta E/E| < 10^{-5}$.

The rectangular barrier configuration (Case 1) employed barrier parameters $V_0 = 4.5$ eV and width $a = 1.0$ nm, yielding transmission coefficient $T = 0.826735$ (82.674\%), reflection coefficient $R = 0.171434$ (17.143\%), and absorbed probability $A = 0.001109$ (0.111\%), satisfying probability conservation with $T + R + A = 0.999278$ (classified as ``excellent''). The Gaussian barrier configuration (Case 2) featured barrier height $V_0 = 4.0$ eV and characteristic width $\sigma_V = 0.8$ nm, producing transmission coefficient $T = 0.786226$ (78.623\%), reflection coefficient $R = 0.209228$ (20.923\%), and absorbed probability $A = 0.001595$ (0.160\%), with total probability sum $T + R + A = 0.997049$ (classified as ``excellent''). Both simulations achieved quality classification thresholds confirming proper implementation of absorbing layers~\cite{Manolopoulos2002derivation,Riss1993calculation}. Initial kinetic energies $E_k^{\text{rect}} = 8.15$ eV and $E_k^{\text{gauss}} = 6.44$ eV both substantially exceed their respective barrier heights, placing the dynamics in the over-barrier tunneling regime.

\begin{figure}[H]
\centering
\includegraphics[width=0.75\textwidth]{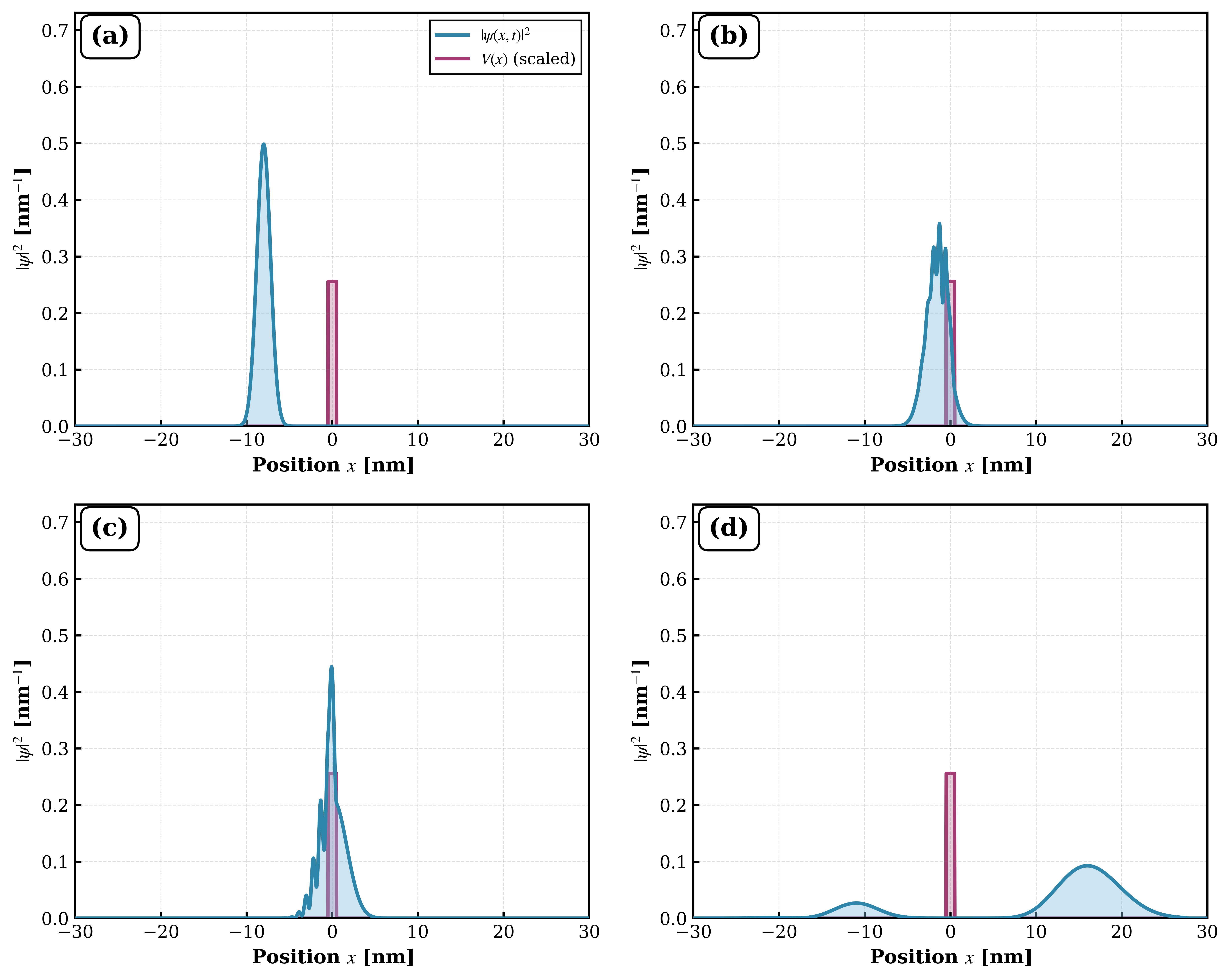}
\caption{Temporal evolution of quantum tunneling through rectangular barrier at $t = 0$, 1.63, 2.08, and 6.0 fs.}
\label{fig:rect_barrier}
\end{figure}

Figure~\ref{fig:rect_barrier} presents the temporal evolution for the rectangular barrier case, illustrating the characteristic stages of quantum tunneling through a sharp potential discontinuity. At $t = 0$ fs (panel a), the initial Gaussian wavepacket exhibits maximum localization at $x_0 = -8.0$ nm with probability density peak $|\psi|^2_{\text{max}} = 0.499$ nm$^{-1}$. The wavepacket propagates ballistically toward the barrier until the pre-collision frame at $t = 1.63$ fs (panel b). Peak interaction occurs at $t = 2.08$ fs (panel c), capturing maximum barrier penetration. The wavepacket undergoes quantum splitting~\cite{Goldberg1967comparison}, with transmitted and reflected components spatially separating. By $t = 6.0$ fs (panel d), complete scattering has occurred with the transmitted packet centered at $\langle x \rangle = 11.6$ nm~\cite{Robinett2000quantum}.

\begin{figure}[H]
\centering
\includegraphics[width=0.75\textwidth]{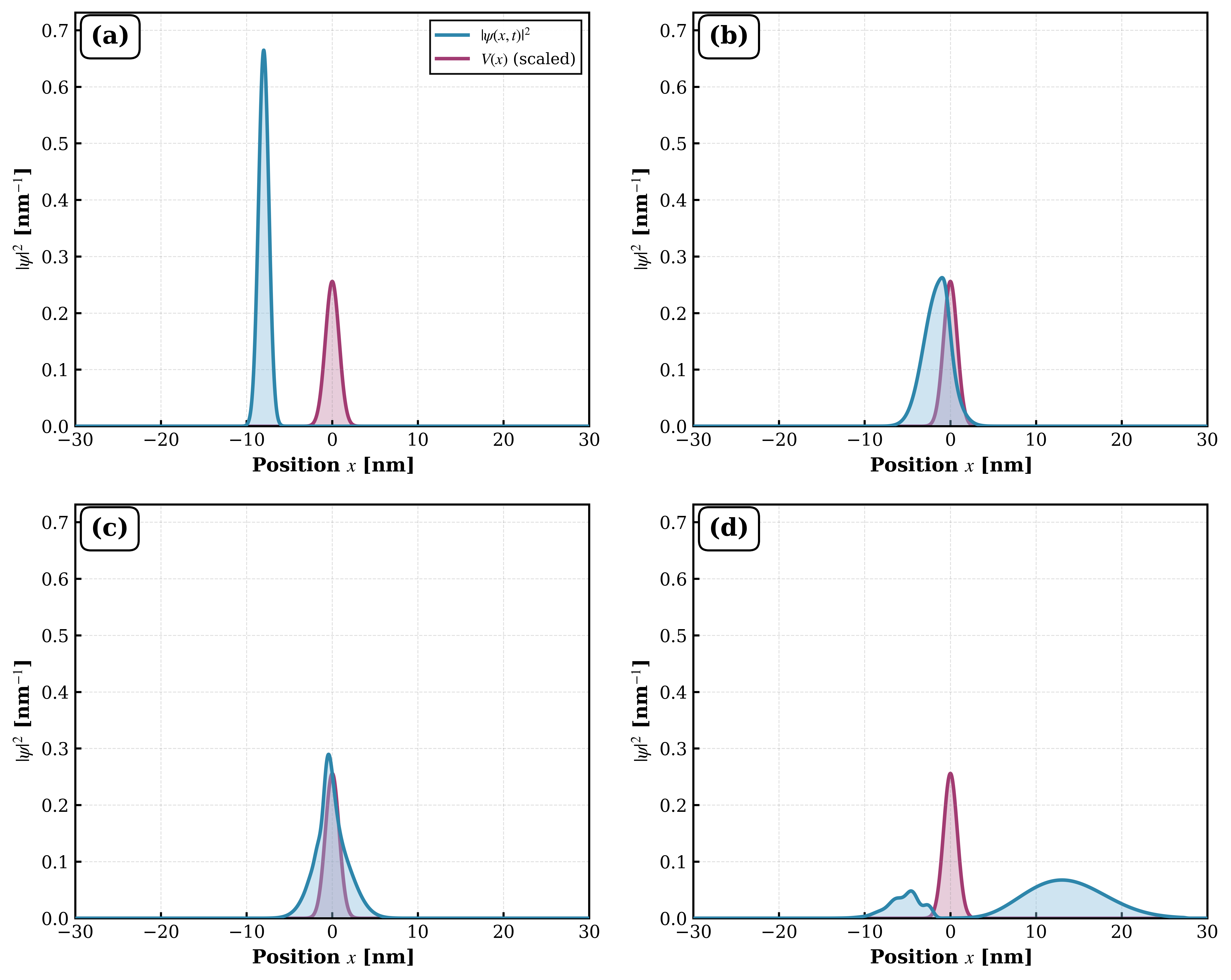}
\caption{Temporal evolution of quantum tunneling through Gaussian barrier at $t = 0$, 1.87, 2.35, and 6.0 fs.}
\label{fig:gauss_barrier}
\end{figure}

The Gaussian barrier case (Figure~\ref{fig:gauss_barrier}) exhibits qualitatively similar temporal progression but with quantitatively distinct scattering characteristics arising from the smooth potential profile. The initial wavepacket at $t = 0$ fs possesses narrower spatial extent ($\Delta x = 0.598$ nm) and higher peak probability density ($|\psi|^2_{\text{max}} = 0.665$ nm$^{-1}$). The peak interaction frame at $t = 2.35$ fs (panel c) shows maximum barrier penetration with probability density peak positioned slightly left of center, suggesting asymmetric tunneling dynamics influenced by the Gaussian profile's curvature~\cite{Razavy2003quantum}. The final scattered state demonstrates lower transmission coefficient $T = 0.786$ compared to the rectangular case ($T = 0.827$).

\begin{figure}[H]
\centering
\includegraphics[width=0.75\textwidth]{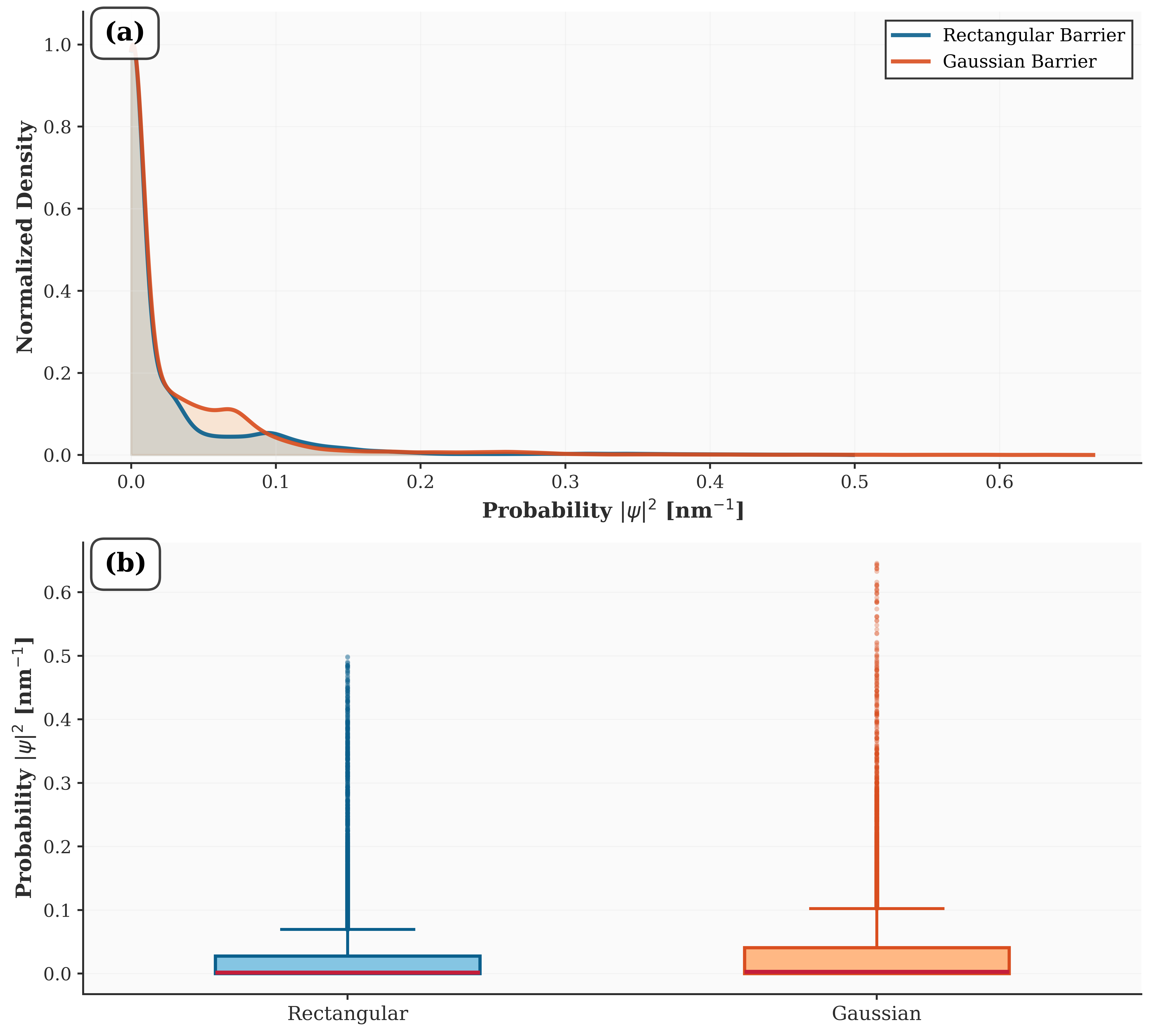}
\caption{Statistical comparison of probability density distributions between rectangular (blue) and Gaussian (red) barriers.}
\label{fig:prob_comparison}
\end{figure}

Comparative analysis of probability distributions across the entire spatial-temporal domain reveals statistically significant differences between barrier geometries (Figure~\ref{fig:prob_comparison} and Table~\ref{tab:statistical_summary}). The kernel density estimation~\cite{Scott1979optimal} demonstrates that both distributions exhibit pronounced peaks near zero probability density. The Gaussian barrier shows higher median probability density ($2.75 \times 10^{-3}$ nm$^{-1}$ versus $1.71 \times 10^{-3}$ nm$^{-1}$) and larger interquartile range, indicating greater variability despite similar mean values (rectangular: $2.91 \times 10^{-2}$ nm$^{-1}$, Gaussian: $3.16 \times 10^{-2}$ nm$^{-1}$, representing $8.7\%$ relative difference). Statistical analysis employed stratified sampling~\cite{Cochran1977sampling} maintaining $10^5$ representative points per distribution.

Information-theoretic analysis provides rigorous quantification of distributional differences. Shannon entropy values $H_{\text{rect}} = 3.054$ bits and $H_{\text{gauss}} = 3.175$ bits reveal the Gaussian barrier produces $3.8\%$ higher entropy. The JS divergence $D_{\text{JS}} = 0.0171$ bits indicates the distributions are remarkably similar~\cite{Lin1991divergence,Endres2003new}. The Kullback-Leibler divergence exhibits notable asymmetry: $D_{\text{KL}}(\text{Rect}||\text{Gauss}) = 0.0635$ bits versus $D_{\text{KL}}(\text{Gauss}||\text{Rect}) = 0.1345$ bits~\cite{Kullback1951information}. The Kolmogorov-Smirnov test strongly rejects distributional equality ($D_{\text{KS}} = 0.0626$, $p < 0.001$), as do five of seven tests, yet the Mann-Whitney $U$ test fails to reject median equality ($p = 0.078$)~\cite{Mann1947test,Kruskal1952use}. Cliff's delta $\delta = -0.018$ indicates negligible practical significance~\cite{Cliff1993dominance,Romano2006appropriate}.

\begin{figure}[H]
\centering
\includegraphics[width=0.75\textwidth]{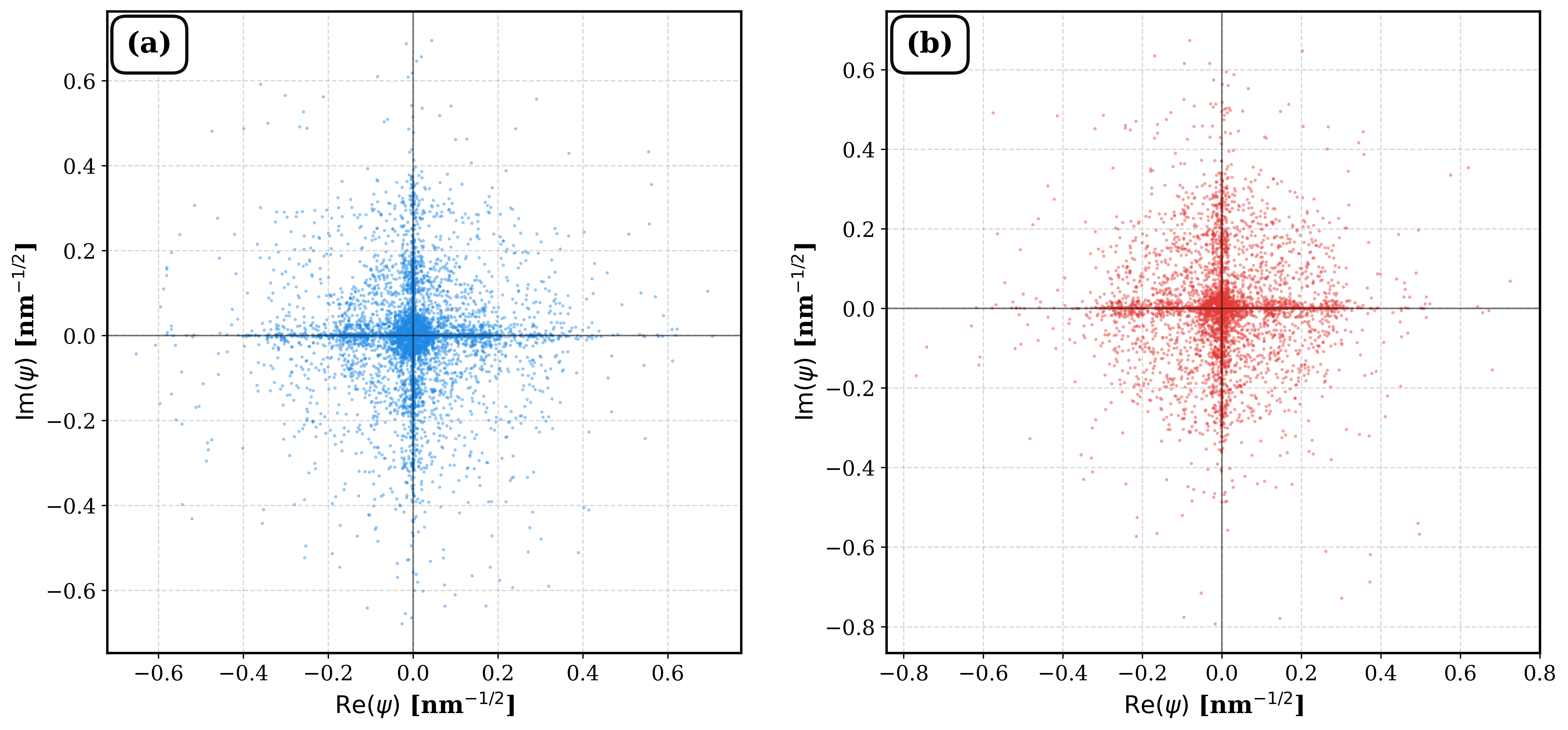}
\caption{Phase space distributions in $(\psi_{\text{Re}}, \psi_{\text{Im}})$ coordinates for (a) rectangular and (b) Gaussian barriers.}
\label{fig:phase_space}
\end{figure}

Phase space analysis in $(\psi_{\text{Re}}, \psi_{\text{Im}})$ coordinates (Figure~\ref{fig:phase_space}) reveals the complex wavefunction's intrinsic structure. Both barrier configurations exhibit approximately circular distributions centered at the origin, characteristic of incoherent superpositions with near-zero correlation. Anisotropy analysis via covariance matrix eigenvalue decomposition~\cite{Jolliffe2002principal} confirms near-perfect isotropy: $\mathcal{A} < 0.001$ for both cases. The effective phase space area reveals the rectangular barrier occupies $17.2\%$ larger area~\cite{Barber1996quickhull}. Phase coherence indices $\text{PCI} < 0.001$ for both barriers~\cite{Lachaux1999measuring} confirm complete decoherence when averaging over the entire spatial-temporal domain.

\begin{table}[H]
\centering
\caption{Statistical comparison of probability density distributions and phase space structures between rectangular and Gaussian barrier configurations.}
\label{tab:statistical_summary}
\small
\begin{tabular}{lcccc}
\hline
\textbf{Measure} & \textbf{Rectangular} & \textbf{Gaussian} & \textbf{Difference} & \textbf{Interpretation} \\
\hline
\multicolumn{5}{l}{\textit{Descriptive Statistics (Probability Density)}} \\
$\mu$ [nm$^{-1}$] & $2.91 \times 10^{-2}$ & $3.16 \times 10^{-2}$ & $2.54 \times 10^{-3}$ & $+8.7\%$ \\
$\sigma$ [nm$^{-1}$] & $6.18 \times 10^{-2}$ & $6.37 \times 10^{-2}$ & $1.87 \times 10^{-3}$ & $+3.0\%$ \\
Median [nm$^{-1}$] & $1.71 \times 10^{-3}$ & $2.75 \times 10^{-3}$ & $1.04 \times 10^{-3}$ & $+60.5\%$ \\
$\gamma_1$ & 3.647 & 4.115 & 0.468 & Gaussian more skewed \\
$\gamma_2$ & 16.41 & 23.49 & 7.08 & Gaussian heavier tails \\
\hline
\multicolumn{5}{l}{\textit{Information-Theoretic Measures}} \\
$H$ [bits] & 3.054 & 3.175 & 0.121 & Gaussian more dispersed \\
$D_{\text{JS}}$ [bits] & \multicolumn{2}{c}{0.0171} & -- & Very similar \\
$H_{\text{PS}}$ [bits] & 4.381 & 3.941 & 0.441 & Rectangular more complex \\
\hline
\multicolumn{5}{l}{\textit{Hypothesis Testing}} \\
$D_{\text{KS}}$ & \multicolumn{2}{c}{0.0626} & $p < 0.001$ & Significant \\
Cliff's $\delta$ & \multicolumn{2}{c}{$-0.018$} & -- & Negligible effect \\
\hline
\multicolumn{5}{l}{\textit{Phase Space Structure}} \\
PCI & 0.000 & 0.000 & 0.000 & No coherence \\
$\mathcal{A}$ & 0.000 & 0.001 & 0.001 & Nearly isotropic \\
$\mathcal{A}_{\text{eff}}$ [nm$^{-1}$] & $3.24 \times 10^{-1}$ & $2.76 \times 10^{-1}$ & $4.74 \times 10^{-2}$ & Rectangular $17.2\%$ larger \\
\hline
\end{tabular}
\end{table}

The observed transmission coefficients $T_{\text{rect}} = 0.827$ and $T_{\text{gauss}} = 0.786$ differ by approximately 5.2\%. The rectangular barrier's higher transmission appears counterintuitive given its sharper potential discontinuities~\cite{Griffiths2018quantum}. However, this result aligns with previous observations that smooth potentials can exhibit enhanced reflection in certain energy regimes due to adiabatic following effects~\cite{Razavy2003quantum}. Both systems operate in the over-barrier regime, yet quantum reflection persists due to wavefunction interference at barrier interfaces~\cite{Landau1977quantum}. The absorbed probabilities $A < 0.002$ remain negligible, confirming that CAP layers~\cite{Manolopoulos2002derivation,Riss1993calculation} successfully eliminate boundary reflections.

Several caveats merit acknowledgment. The analysis considers only two specific barrier configurations; systematic parameter variation would be necessary for general scaling relationships. The simulations employ one-dimensional reduction, neglecting transverse degrees of freedom~\cite{Garcia-Moliner1976theory}. Environmental decoherence effects~\cite{Zurek2003decoherence} are absent. The negligible effect size ($|\delta| = 0.018$) suggests that practical differences may be modest for this particular over-barrier scenario. Sub-barrier tunneling scenarios ($E_k < V_0$) might exhibit more pronounced geometric effects~\cite{Razavy2003quantum}. These results demonstrate that the \texttt{1d-qt-ideal-solver} successfully resolves quantum tunneling dynamics with high numerical fidelity, establishing a validated computational framework for future investigations~\cite{Ricco1984physics,Mizuta1995physics,Grossmann1991coherent,Joos2003decoherence}.

\section{Conclusions}

We have developed and rigorously validated \texttt{1d-qt-ideal-solver}, a production-grade Python library for simulating quantum tunneling dynamics through arbitrary one-dimensional potential barriers under \emph{idealized} conditions. The library name explicitly reflects its scope: this solver implements coherent unitary evolution without dissipation, environmental coupling, or many-body interactions, representing an intentionally simplified model valuable for pedagogical exploration and qualitative insight rather than quantitative prediction of experimental observables. The split-operator method with CAPs achieves second-order temporal accuracy while maintaining exceptional numerical stability, as evidenced by energy conservation $|\Delta E/E| < 10^{-5}$ and probability conservation $|T+R+A-1| < 10^{-3}$ over femtosecond-scale propagation times. Comparative analysis of rectangular versus Gaussian barriers in the over-barrier regime ($E_k > V_0$) reveals a counterintuitive 5.2\% transmission enhancement for the rectangular geometry ($T_{\text{rect}} = 0.827$ versus $T_{\text{gauss}} = 0.786$), which we tentatively attribute to reduced adiabatic following effects in sharp potential discontinuities. While comprehensive statistical analysis employing information-theoretic measures and non-parametric hypothesis tests confirms statistically significant distributional differences between barrier configurations, the negligible effect size (Cliff's $\delta = -0.018$) and low JS divergence ($D_{\text{JS}} = 0.0171$ bits) suggest that practical differences remain modest for this particular energy regime---though sub-barrier tunneling scenarios ($E_k < V_0$), where transmission becomes exponentially sensitive to barrier geometry, may exhibit substantially more pronounced effects warranting future investigation. Several critical limitations must be acknowledged: (1) our analysis considers only idealized coherent evolution, neglecting environmental decoherence, (2) the one-dimensional reduction neglects transverse degrees of freedom and lateral confinement effects crucial to realistic nanoscale devices, (3) thermal effects, phonon coupling, and electron-electron interactions are entirely absent, (4) the single-particle approximation breaks down for many experimentally relevant scenarios, and (5) validation was performed only for two specific barrier configurations. Nevertheless, the solver's modular architecture, \texttt{numba}-accelerated performance (50$\times$ speedup achieving $\sim$12-second wall times), and comprehensive validation establish a robust computational framework for educational applications and exploratory investigations of fundamental tunneling phenomena. The open-source \texttt{1d-qt-ideal-solver} library, distributed under the permissive MIT License, provides the computational physics community with an immediately deployable tool for teaching quantum mechanics, rapid prototyping of potential profiles, and preliminary exploration of tunneling dynamics.

\section*{Acknowledgements}

The authors used Claude Sonnet 5 strictly for language editing and assume full responsibility for all manuscript content and remaining imperfections. This study was funded by the ITB 3P Research Program 2026 (Talenta Unggul Scheme) through the Directorate of Research and Innovation, Bandung Institute of Technology (Project ID: DRI.PN-6-64-2026).

\section*{Author Contributions}

S.H.S.H.: Conceptualization; Formal analysis; Investigation; Methodology; Software; Validation; Visualization; Funding acquisition; Writing -- original draft; Project administration. S.F.B.: Conceptualization; Funding acquisition; Supervision; Writing -- review \& editing. S.N.K.: Formal analysis; Writing -- original draft. R.S.: Funding acquisition; Supervision; Writing -- review \& editing. N.J.T.: Supervision; Writing -- review \& editing. I.P.A.: Funding acquisition; Supervision; Writing -- review \& editing. D.E.I:  Funding acquisition; Supervision; Writing -- review \& editing. All authors reviewed and approved the final version of the manuscript.

\section*{Open Research}

The \texttt{1d-qt-ideal-solver} library is freely available as open-source software under the MIT License. The complete source code can be accessed through the GitHub repository at \url{https://github.com/sandyherho/1d-qt-ideal-solver}. The library can be installed directly via the Python Package Index (PyPI) at \url{https://pypi.org/project/1d-qt-ideal-solver/}. All Python scripts used for data analysis, statistical computations, and figure generation presented in this manuscript are available in a separate GitHub repository at \url{https://github.com/sandyherho/suppl_1d-qt}. The data supporting this publication are permanently archived and publicly accessible through the Open Science Framework (OSF) repository at \url{https://doi.org/10.17605/OSF.IO/RVDQ2}. All repositories are distributed under the permissive MIT License.

\end{document}